# DESIGN AND ANALYSIS OF SD_DWCA – A MOBILITY BASED CLUSTERING OF HOMOGENEOUS MANETs

T.N. Janakiraman[1] and A. Senthil Thilak[2]

[1, 2]Department of Mathematics, National Institute of Technology, Tiruchirapalli-620015, Tamil Nadu, India

[1]`janaki@nitt.edu, tnjraman2000@yahoo.com`
[2]`asthilak23@gmail.com`

## ABSTRACT

*This paper deals with the design and analysis of the distributed weighted clustering algorithm SD_DWCA proposed for homogeneous mobile ad hoc networks. It is a connectivity, mobility and energy based clustering algorithm which is suitable for scalable ad hoc networks. The algorithm uses a new graph parameter called strong degree defined based on the quality of neighbours of a node. The parameters are so chosen to ensure high connectivity, cluster stability and energy efficient communication among nodes of high dynamic nature. This paper also includes the experimental results of the algorithm implemented using the network simulator NS2. The experimental results show that the algorithm is suitable for high speed networks and generate stable clusters with less maintenance overhead.*

## KEYWORDS

*Mobile ad hoc networks, SD_DWCA, Connectivity, Strong degree, Cluster Stability*

## 1. INTRODUCTION

Mobile Ad hoc Networks [MANETs], which are otherwise called as multi-hop networks or peer-to-peer networks are those formed by a series of autonomous mobile hosts interlinked by means of bandwidth-constrained wireless links with limited battery power. Ad hoc networks, as their name indicates, are unpredictable with frequently changing topology. Each node has a circular transmission range and those nodes which lie within this range alone can communicate with this node. The transmission range of all the nodes can either be uniform or may vary and the nodes may either be of similar or of dissimilar nature. If the transmission range of all the nodes is uniform or the nodes are of same nature, i.e., they are of same architecture then the network is termed as a Homogeneous network. Otherwise, it is referred to as a heterogeneous network. The management and control functions of the network are distributed among the nodes in the entire network. As the network is highly decentralized, all network activities including topology discovery, keeping track of topological changes due to mobility, transmitting and maintaining routing information and efficient usage of battery power must be executed and monitored by the nodes themselves. Since these networks are highly dynamic and ad hoc in nature, executing these control functions becomes a bottle-neck. In order to perform this management and control operations and to communicate messages among nodes having no permanent bonding, it is essential to set up a virtual backbone and this is accomplished by the process of clustering.

Clustering is the process of grouping nodes based on certain strategies and the above mentioned functions can either be distributed among the nodes in each cluster or performed by the leader nodes/Cluster-heads of each cluster rather than distributing throughout the network. This reduces, to a great extent, the information exchange between the network nodes and the information to be maintained by each node thereby reducing the overheads incurred. Several algorithms are available for clustering of mobile ad hoc networks [1], [12], [15], [18], [21], [24]. Particularly, classified as mobility-only-based algorithms [5], [13], [16], power-only-based [2], [18], [19] and combination-based algorithms [7-9], [18], [21], [22]. The algorithm discussed in this paper is a combination-based algorithm, proposed in [23]. In our algorithm, a new graph parameter called strong degree is defined and in addition, we consider two other parameters, namely, mobility metric [introduced by Xing et al. in 20] and battery power.

The rest of this paper is organized as follows. The basic definitions and graph theoretic terminologies used in this algorithm are given in section 2 and section 3 outlines a review of the existing clustering algorithms. Section 4 includes the basic idea which led to the development of the algorithm and the objectives of the algorithm. Section 5 gives the description and section 6 the complexity analysis of the algorithm. Section 7 says how the clusters can be ranked based on their cluster members. Section 8 includes the simulation study and finally, Section 9 concludes the paper.

## 2. BASIC DEFINITIONS AND TERMINOLOGIES

In this paper, the mobile ad hoc network taken under consideration is assumed to be homogeneous, i.e., each node has uniform transmission power. Hence, from now on, by a network, we mean a Homogeneous Mobile Ad Hoc Network unless otherwise specified. In a general wireless network, to identify the neighboring nodes, each node broadcasts a "Hello" message containing information like its position in the terrain, node id etc. Those nodes which are within the transmission range of the transmitting node can receive the message and they either send an acknowledgement to the sender or simply store the information received from the sender to identify and record details about their neighbors. These recorded data, in turn can be utilized for clustering and/or routing. As the nodes are mobile, each node periodically should send and receive the hello messages, to keep track of their neighbor info. Thus, the neighbor set is periodically updated for each node. This communication strategy can in general be modelled by using an undirected graph as follows.

### 2.1. Graph Theory vs. MANET

A graph $G$ is defined to be an ordered pair *(V, E)*, where *V* is a non-empty set of nodes and *E* denotes the set of edges/links between different pairs of nodes in *V*. Any set of devices which are capable of interacting with each other can be modeled by using graphs. To represent any given network using a graph, say *G*, the set of all nodes in the network is taken as the node set *V*, where two nodes are made adjacent, if the corresponding two nodes are within the transmission range of each other, i.e., each is in the neighbour set of the other. The graph *G* thus obtained is referred to as the *underlying graph or the network graph.*

If u and v are any two nodes in *G*, then *d(u, v)* denotes the least number of hops to move from u to v and vice versa and is referred to as the *Hop-distance* between u and v and *ed(u, v)* denotes the Euclidean distance between u and v. Thus, in a homogeneous network, for a given transmission range r, two nodes u and v can communicate with each other if and only if they are at an Euclidean distance less than or equal to r, i.e., ed(u, v) ≤ r . For a given node u, *N(u)* denotes the set of neighbors of u, i.e, *N(u)* is the set of those nodes which are 1-hop away from u and its cardinality is defined as the *degree* of u, which is denoted by *deg(u)*. The hop-distance between u and its farthest node in G is called the *eccentricity of u* in G and is denoted by *ecc(u)*, i.e., $ecc(u) = \max_{v \in V(G)} \{d(u, v)\}$. The minimum and maximum eccentricities of *G* are defined respectively as *radius* [*r(G)*] and *diameter* [*d(G)*] of G. A subset *S* of the node set *V(G)* of a graph G is said to be a *dominating set* of *G*, if each of the nodes in the network is either in *S* or is adjacent to some node in *S*.

## 2.2. Categorization of Neighbors of a Node

For a given node u (transmitting node), the nodes which are closer to u will receive stronger signals and those nodes which are far apart from u will get weaker signals. Based on this notion, we classify the neighbors of a transmitting node as follows [23], [24]:

    i.  Strong neighbor
    ii.  Weak neighbor
    iii. Medium neighbor

**Strong neighbor:** A node v is said to be a *strong neighbor* of a node u, if the Euclidean distance between u and v is less than or equal to r/2. i.e., $0 \le ed(u, v) \le r/2$.

**Medium neighbor:** A node v is said to be a *medium neighbor* of a node u, if $r/2 < ed(u, v) \le 3r/4$.

**Weak neighbor:** A node v is said to be a weak neighbor of a node u, if $3r/4 < ed(u, v) \le r$.

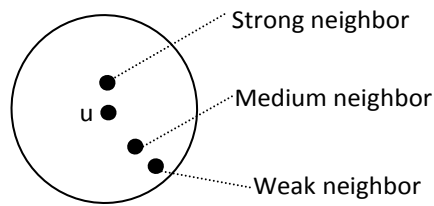

Figure 1.  Three types of neighbors of node u

# 3. PRIOR WORK

As mentioned earlier, because of the ad hoc and dynamic nature of the networks under consideration, it is essential to set up a virtual backbone so as to facilitate effective communication. Further, due to frequent changes in the topological information, it is very difficult to maintain the neighbor and routing table for the entire set of nodes in the network. Therefore, to provide an optimal cost effective communication, the amount of information maintained by each node (which is essential for controlling and routing) is reduced to a certain extent by constructing a hierarchical organization of the network. The organization should be in such a way that it is retained as long as possible and changes less frequently. This hierarchical organization can be provided by the process of clustering. Several clustering procedures are available in the literature for clustering of MANETs. They are categorized in different ways. One such classification is the Cluster-Head (CH) based and Non-Cluster-Head (NCH) based clustering algorithms [10], [21] depending on whether a special mobile node named cluster-head, is required for the formations of clusters or not. The algorithm discussed in this paper comes under the category of CH based clustering algorithms. So, few of the related CH based algorithms are discussed below.

**Lowest ID Heuristic (LID).** In the Lowest ID (LID) heuristic proposed by Baker and Ephremides [3, 4], every node in the network is assigned a unique identifier (ID) and the node IDs are broadcast to the respective neighbors. Each node compares its ID with those of its neighbors and it declares itself as a Cluster-Head, if its ID is the lowest among all its neighbors. The clusters are formed with such nodes as Cluster-Heads and their neighbors as respective cluster members. If a node can hear two or more cluster-heads it is designated as a cluster-gateway. The major drawback of the algorithm is that when a highly mobile node is chosen as cluster-head, being the node with smallest id among its neighbors, it causes inconvenient re-clustering and undesired cluster-head changes. Secondly, since this algorithm is inclined to choose nodes having smaller IDs as cluster-heads, the smaller ID nodes suffer from battery drainage leading to short lifespan. Further, the id is not assigned based on any practical constraint and hence, the election of cluster-heads is not meaningful.

**Highest Degree Heuristic (HD).** The highest degree (HD) heuristic is based on the connectivity between a node and its neighbors and was proposed by Gerla et al. [11, 17]. This connectivity value of a node is broadcast to all its neighbors and if a node has higher connectivity value than its neighbors, then it is elected as a cluster-head and a cluster is formed with the elected cluster-head and its neighbors. Any tie in cluster-head election is broken by node IDs. Cluster-gateways are formed with the nodes capable of hearing two or more cluster-heads. Since this algorithm doesn't restrict the number of nodes that can be handled by a cluster-head, the heads run out of power very quickly.

**Mobility-only-based Clustering Algorithms.** In MOBIC given by Basu et al. [5], the relative mobility metric is introduced, in order to form stable clusters and is defined as the logarithm of the ratio of two successive received signal strengths and an aggregate relative mobility metric $M$ is computed for each node by taking the variance (with respect to zero) of the entire set of pairwise relative mobility values of the node and each of its neighbors. Then it follows the same procedure as in LID heuristic to form clusters with node ID in LID replaced by the value M here. Hence, it has all the drawbacks of LID. The same problem is met with the MobDHop clustering algorithm given by Er and Seah [13].

**Power-only-based Clustering Algorithms.** In general, for any cluster-head election procedure, a node with higher battery power is considered to be a better candidate to play the key role of a cluster head. In that way, the algorithms given in [18], [19] consider battery power as the only system parameter for electing cluster-heads. But, as the mobility of nodes is not considered in the election, the possibility of re-clustering is still high when elected cluster-heads have high mobility.

**Combination-based Clustering Algorithms.** Chatterjee et al. [7, 8] and Choi and Woo[9] have given clustering algorithms using multiple-metrics. In each of these algorithms, they considered mobility and battery power as two of the important metrics together with other parameters. In all these algorithms, the average speed is computed by considering the (x, y) positions of the nodes at various instants (from initial time t = 1 to t = T, where T is the time at which the algorithm is executed or weight is computed) and is used as the mobility metric. The (x, y) positions are obtained using GPS, the accuracy of which is not ideal for fine computing and the operations of which could drain the limited battery power of the node quickly. Further, regarding power metric, Chatterjee et al. [7, 8] used the cumulative time period for which a node acts as a cluster-head for computing battery power. But, this cannot accurately reflect the current level of battery power because a busy node (i.e., a node which might have lost much of its energy in just transferring or forwarding the control packets) may almost run out of power without being a cluster-head [20]. Choi and Woo [9] used consumed battery power as a metric for computing battery power. Thus, integrating these two (mobility and power) parameters together with the other parameters, node weight is defined for each node in those algorithms and the clustering procedure is similar to that of HD heuristic except that here node weight is used in the place of node degree.

## 4. MOTIVATION AND OBJECTIVE OF THE ALGORITHM

In the algorithm analyzed in this paper, we consider three important parameters, namely, *battery power, mobility measure and strong degree*. Though there are several algorithms available in the literature, starting from the highest degree heuristic, almost all the above mentioned algorithms use the degree of a node as an important parameter and priorities were given to nodes having highest degree. While choosing the nodes with highest degree for acting as cluster-head, all the strong, medium and weak neighbours of a node contribute for the highest degree of that node. But, there may be some nodes, say, u and v such that the number of strong neighbours of u is less than that of v and the number of medium (and weak) neighbours of u is greater than that of v with deg(u) > deg(v). Thus, as per the strategies used in the existing algorithms, on comparing u and v, u becomes eligible for CH election, even though it has lesser number of strong neighbours than v. But, at a later instant, due to mobility, the node u may lose its eligibility of remaining as a CH. This makes the cluster structure less stable and the cluster members attached to that cluster have to be re-affiliated or sometimes re-clustering has to be done.

This is because, the nodes which are at the rim of the circular transmission range, (weak neighbors) of the cluster-head and the nodes which are considerably far away from the cluster-head (medium neighbors) are more likely to move away from the cluster-head, compared to strong neighbors. This factor highly affects the stability of the clusters generated and this leads to frequent cluster-head updates, re-affiliation, re-clustering etc. and hence cluster-maintenance cost is also increased. This problem motivated us to start defining this concept. Considering such a situation, we've defined a new degree called strong degree and this was included as one of the main parameters in SD_DWCA [23]. Apart from this, we also include the mobility of the nodes and the battery power for CH election. For, if a CH is allowed to continue its status for a longer period, then it will lead to battery drainage and hence we should continue with re-affiliation or re-clustering. Hence, to avoid excessive usage of a CH, battery power is considered for selecting cluster heads. With these three parameters, the algorithm is developed with the following objectives.

1. The network nodes are partitioned into different groups of various sizes, so as to form a hierarchical organization of the network.
2. The clusters have to be stable as long as possible, but, without excessive battery drainage.
3. The cluster formation and maintenance overhead should be minimized.
4. The cluster-heads should not be overloaded.
5. During the cluster set up and maintenance phase, the load should be distributed among all the nodes in the network.
6. Re-affiliations should be minimized.
7. Re-clustering should be avoided as much as possible. At times of necessity, re-affiliations are allowed instead of re-clustering to reduce the cost of cluster maintenance.
8. The algorithm should overcome the problem of scalability.
9. The generated clusters should facilitate hierarchical routing.

## 5. STRONG DEGREE BASED DISTRIBUTED WEIGHTED CLUSTERING ALGORITHM (SD_DWCA)

### 5.1 Metrics

**5.1.1. Strong Neighbour Set -** Let $G = (V, E)$ be the underlying network graph and $u \in V(G)$. The *Strong neighbour set* of u is defined to be the set of all strong neighbours of u and is denoted by *SN(u)*. i.e., $SN(u) = \{v \in N(u): 0 \le ed(u, v) \le r/2\}$

**5.1.2. Strong Degree -** The *Strong degree* of a node $u \in V(G)$ is the cardinality of the set *SN(u)* and is denoted by $d_{sn}(u)$.

In the same way, the *medium neighbour set (denoted by MN(u))* and *weak neighbour set (denoted by WN(u))* are defined using the respective definitions given in section 2.2. The *medium degree* of a node $u \in V(G)$ is the cardinality of the set *MN(u)* and is denoted by $d_{mn}(u)$ and the *weak degree* of a node $u \in V(G)$ is the cardinality of the set *WN(u)* and is denoted by $d_{wn}(u)$.

In general, the degree of a node is the sum of the number of strong neighbours, medium neighbors and weak neighbours, i.e., $deg(u) = |SN(u)| + |MN(u)| + |WN(u)| = d_{sn}(u) + d_{mn}(u) + d_{wn}(u)$.

### 5.1.3 Mobility metric:

Here, we adopt the mobility measure defined by Xing et al. [21]. For each node u in the network, after receiving two successive hello messages from its 1-hop neighbors, the relative mobility metric with respect to each of its 1-hop neighbors is computed using the formula (1). If node v is a neighbor of node u, then the relative mobility metric of u with respect to v is denoted by $R_v(u)$, and is defined as follows:

$$R_v(u) = \frac{r_{1v}}{r_{2v}} \qquad\qquad \text{------------------------(1)}$$

Here, $r_{1v}$ and $r_{2v}$ denote the two successive signal strengths received by u from v. This received signal strength can be read from RSS indicator. The signal strengths received $r_{1v}$ and $r_{2v}$ may be like either $r_{1v} < r_{2v}$ or $r_{1v} = r_{2v}$ or $r_{1v} > r_{2v}$. Thus, we have the following cases.

**Case i:** $R_v(u) < 1$
    If $R_v(u) < 1$, then it indicates that the nodes u and v move close to each other.
**Case ii:** $R_v(u) > 1$
    If $R_v(u) > 1$, then it indicates that the nodes u and v move away from each other.
**Case iii:** $R_v(u) = 1$
    If $R_v(u) = 1$, then it indicates that either the nodes u and v do not move at all or they move with the same speed in the same direction.

In this way, the relative mobility value of a node is computed with respect to each of its neighbors and the root mean square deviation of these values taken from the value 1 (as given in equation (2)) is used as the mobility measure of u and is denoted by M(u).

$$M(u) = \sqrt{\frac{\sum\limits_{v \in N(u)} (R_v(u) - \overline{R}(u))^2}{deg(u)}} \,, \qquad\qquad \text{--------------------------(2)}$$

where $\overline{R}(u) = 1$.

In the cluster-head election procedure, to maintain the stability of elected cluster-heads, the cluster-heads are expected not to move away from their cluster members. Hence, the value $R_v(u)$ of each node u with respect to each of its neighbours is preferred to be less than or equal to 1 and therefore $\overline{R}(u) = 1$, instead of actual mean. Suppose, some neighbour of node u has sent more than two periodical hello messages, but u didn't receive two of them successively or u received only one of them, then that neighbour is excluded in the weight calculation of node u. This is because, that particular neighbour of u might have moved away from u, i.e., might be a weak neighbour of u or there might be signal attenuation.

**5.1.4. Power Metric:**

To consider the battery level of each mobile node, and to ensure that the cluster-heads are not excessively utilized, i.e., used until they get completely drained out, we use a metric denoted by RE(u), which denotes the residual energy/remaining battery power of node u. If IE(u) denotes the initial energy assigned to node u and CE(u) denotes the consumed battery power of u, then, RE(u) = IE(u) − CE(u). To compute this value, each node records its remaining battery power after sending and receiving every message.

## 5.2 Node Weight

To elect CHs, each node is initially assigned a weight computed using formula (3).

$$W(u) = \alpha_1\, d_{sn}(u) + \alpha_2[1/M(u)] + \alpha_3\, RE(u) \qquad \text{----------------------(3)}$$

Here, the constants $\alpha_1$, $\alpha_2$ and $\alpha_3$ denote the weighing factors. In almost all types of applications, it is natural to give equal weightage to all the three parameters which we've considered. So, we choose the values of the three weighing factors as 1/3, so that their sum is unity. If in some application, any of these parameters have to be given special importance then the weighing factors can be chosen accordingly.

Here, the first parameter, strong degree is chosen to maintain stability of cluster-heads and thereby, the stability of the generated clusters, the second parameter is chosen to make the algorithm to be adaptive to mobility and the third to choose a node which can afford to play the key role of a cluster-head for a longer period, compared to others.

The algorithm includes two phases, namely, cluster formation and cluster maintenance as explained below.

## 5.3 Cluster Formation

At any instant t, Let $G_t$ denote the graph corresponding to the underlying topology of the nodes at that particular instant and $V(G_t)$ denote the node set of $G_t$. It is assumed that all nodes send and receive data with equal transmission range and the nodes are moving with velocity ranging in the interval $[0, V_{max}]$ and they are free to move in any direction. The status of each node is initially set to "UNKNOWN".

**5.3.1. Initial Cluster Formation.**

*Step 1:* Each node periodically sends and receives hello messages. The hello messages are sent with a predefined broadcast interval (BI).

*Step 2:* While sending a hello message, a node sends its node id, (x, y) position (to categorize the neighbor). Meanwhile, the sender should record its remaining battery power.

*Step 3:* A node, upon receiving a hello message, stores the neighbor info received. Also the received signal strength and the time stamp are recorded.

*Step 4:* As specified, if a node fails to receive two successive hello messages after a neighbor has sent atleast three messages, then the neighbor is excluded in weight computation.

*Step 5:* With the above information, each node computes its own weight after a fixed duration. This time gap is specified to allow the hello messages to be sent and received by each node from all its neighbors.

*Step 6:* After computing the weight value, each node broadcasts this weight value to all its neighbors.

*Step 7:* Then, each node compares its weight with that of its neighbors. If the weight of a node is maximum, then it declares itself as a cluster-head and sends a message "CLUSTER_INFO" together with its node id, to all its neighbors.

Here, each node has an active participation in the cluster formation process and hence the algorithm is referred to as distributed.

*Step 8:* On receiving the message CLUSTER_INFO, if the current status of the receiving node is UNKNOWN, then it becomes a cluster member of the sender and stores the sender's id as its cluster-head id and in turn broadcasts the message to all its neighbors. Otherwise, if it is already a member of another cluster and if the weight of its cluster-head is less than that of the sender, then it changes its affiliation, i.e., changes its cluster-head id as the id of the sender.

*Step 9:* Finally, the clusters generated are obtained as output.

*Step 10:* After making all comparisons, nodes still in the status of "UNKNOWN" are collected separately and are termed as "Critical nodes". The critical nodes are then subjected to an adjustment procedure.

### 5.3.2. Formation of Adjusted Clusters.

Let C denote the set of critical nodes obtained after initial cluster formation. If some (atleast two) of the critical nodes are adjacent, then they are allowed to form a cluster on their own. In this case, among all the neighbors, the one with higher weight is chosen as the Cluster-head and the rest are considered as cluster members. If still there are left out nodes, then they become cluster-heads on their own. This process distinguishes our algorithm from the other existing algorithms. In most of the existing algorithms, if there is any node left uncovered, then either re-clustering takes place [7], [8] or the nodes are themselves declared as cluster-heads [9]. But, in the proposed algorithm, the nodes are grouped as far as possible, thereby reducing the number of clusters generated.

## 5.4 Cluster Maintenance

In general, while dealing with mobile networks, the position (position in the terrain) and status (cluster-head/member/critical node, in our case) of the nodes at the time of cluster formation will change in due course, because of nodes' mobility. Hence, in order to provide a valid hierarchical structuring to facilitate an efficient routing, it is necessary to consider this topology change and discuss the behavior of the clustering procedure in such cases. This leads to the cluster maintenance phase. The change in the initial topology may be caused due to node failure because of battery drainage, addition of a new node into network, link failure, link establishment. Hence, we discuss the behavior of our algorithm in all these four cases separately.

### Case i: Node failure

As per the clustering procedure given in [23], a node failure means either the drainage of the battery power of a node below a fixed threshold value or complete exhaustion of battery power of a node. Generally, the idea of clustering is adopted to properly utilize the resources of the nodes such as battery power, limited bandwidth etc, in data transmission. Hence, the cluster-heads are allowed to play a key role than the non-cluster heads. This leads to higher usage of battery power by cluster-heads rather than the other nodes. So, we assume that only the cluster-heads have a greater chance of getting drained out quickly and we deal with the failure of cluster-heads. To avoid cluster-heads being overloaded or to avoid excessive usage of battery by a subset of the nodes, we fix a threshold value for energy/battery power. During the formation of clusters, each cluster-head periodically checks whether its residual energy is above the threshold value. When it goes below the threshold value, the cluster-head sends a resignation message and all the nodes in the cluster should affiliate themselves to other existing clusters. Hence, each cluster member of such a cluster will send a find_CH message to all its neighboring cluster heads. Any cluster-head which receives this message will in turn send an acknowledgement message to the senders and includes that node in its list of cluster members.

If a sending node receives a cluster-head acknowledgement message from more than one node, then it gets attached to the one having maximum weight. When a node gets completely drained out, it will be isolated from communication. But, the periodical check up of available battery power of the maximum utilized nodes will avoid such a situation as far as possible. This also increases the network life time.

### Case ii: Node addition

There is a chance for a new node entering into the network. In such a case, the newly entering node should attach itself to an existing cluster. Hence, after identifying its neighbors by passing hello messages, it broadcasts the find_CH message and fixes its appropriate status as explained in the previous case.

### Case iii: Link failure

There is also a possibility for the nodes which are already grouped into some cluster to move outside the boundary of the cluster. Then, such a node also sends a find_CH message and gets attached to some existing cluster as in case i.

### Case iv: Link establishment

When there is a possibility for nodes attached to one cluster moving outside its existing cluster boundary, the same will move closer to some other cluster boundary. This will induce new link establishments and also continuous movement of the nodes will also induce frequent link establishments and failures. To handle new link establishments, each cluster-head should periodically update its neighbor table. If a cluster-head finds a new node in its neighbor table, it checks whether its weight is greater than that of the new node. If it is greater, it continues its status and adds the new neighbor as its cluster member. If not, it will check in its neighbor table whether there are many (quantified as atleast two) nodes which are common neighbors of both the existing cluster-head and the newly added neighbor, then the current cluster-head will interchange its role with the newly added neighbor. The nodes in the current cluster, which are non-neighbors of the newly elected cluster-head will get attached to the appropriate neighboring clusters by passing find_CH message.

If an execution of the above clustering procedure yields no critical nodes, then the clustering is said to be a *perfect clustering.* On the other hand, if there are some critical nodes left out and the number of critical nodes can be reduced to zero by the adjustment procedure, then the clustering is said to be a *fairly perfect clustering* [23].

## 6. COMPLEXITY ANALYSIS

To perform the analysis, it is assumed that the continuous run time of the algorithm is divided into discrete time steps. Here, one time step is defined as the time duration for the sending of a message (control packets) by a sender and a complete processing of it by the recipient [6]. The approach adopted in this paper is motivated by the theoretical analysis of DMAC made in [6], [13]. It is also assumed that any message transmitted by a sender is successfully received and processed by all its recipients in one time step. We compute the total overhead (message/time) incurred in three steps as follows: Overhead due to hello protocols, cluster formation and cluster maintenance.

Let $M_H$, $M_{CF}$, $M_{CM}$ denote the message complexities due to hello protocol, cluster formation, cluster maintenance respectively and $T_H$, $T_{CF}$, $T_{CM}$ respectively denote the time complexities due to hello protocol, cluster formation, cluster maintenance and M and T denote the overall message and time complexities respectively. The total message complexity of the algorithm is the sum of $M_H$, $M_{CF}$ and $M_{CM}$. Similarly, T is the total of $T_H$, $T_{CF}$ and $T_{CM}$.

### 6.1 Overhead due to Hello Protocols

Each node periodically broadcasts the hello messages to keep track of its knowledge about its neighbours, at a fixed predefined interval BI. Thus, the frequency of hello messages broadcast by a node is T/BI, where T is the total running time of the algorithm, until which each node has to maintain its neighbour data. Hence, on the whole, N*(T/BI) hello messages are to be transmitted, for the entire set of nodes to maintain their neighbour data and so the message complexity due to hello protocols is $M_H = \Theta(N)$, as (T/BI) is a fixed constant, predefined before execution. By our assumption, it takes one time step for the transmission of hello messages by each node and successful reception of it by all its neighbours. Hence, totally N time steps are required for successful transmission and reception. Thus, $T_H = \Theta(N)$.

### 6.2 Overhead due to Cluster Formation

Cluster formation is done by invoking the two procedures in section 5.3.1 and 5.3.2. Suppose that there are $N_c$ clusters and $C_r$ critical nodes before the execution of adjustment procedure. Then, the total number of CMs in the network will be $N - N_c - C_r$.

In the process of cluster formation, initially, each node computes its weight value in constant time and broadcasts this weight value to all its neighbours, so that $M_1 = N * \Theta(1)$. The respective neighbours store this received weight info in their neighbour table for further comparison in one time step, so that $T_1 = N * \Theta(1)$.

Upon receiving this weight info, each node compares its weight with that of its neighbours and decides whether it is maximum. If maximum, it broadcasts a *CLUSTER_INFO* message to all its neighbours. But, this message broadcast is done only by the CHs. Therefore, totally, if $N_c$ is the total number of clusters generated, then $N_c * \Theta(1)$ *CLUSTER_INFO* messages are transmitted, i.e., $M_2 = N_c * \Theta(1)$. Each node upon receiving this *CLUSTER_INFO* decides its role as whether to become a CM of the sender (if its current status is *UNKNOWN*) or change its affiliation (if already a CM) or to retain/change its status based on weight comparison (in case, it is already a CH). This takes one time step for each node, i.e., $T_2 = N_c * \Theta(1)$. After this decision making process, each CM broadcasts its CH id to all its neighbours and the neighbours store this info for further processing. Hence, $M_3 = (N - N_c - C_r) * \Theta(1) = T_3$.

Regarding the adjustment procedure, the critical nodes will look into their neighbor table for nodes with *UNKNOWN* status. If such neighbours are available, then the adjusted clusters are formed based on weight comparison. The nodes which declare themselves as CHs will also send *CLUSTER_INFO* messages to their neighbors. In the worst case, all the critical nodes can become as CHs so that $C_r$ messages are transmitted and $M_4 = C_r * \Theta(1) = T_4$.

Therefore, the total message and time complexities due to cluster formation are given by,

$$M_{CF} = M_1 + M_2 + M_3 + M_4$$
$$= N * \Theta(1) + N_c * \Theta(1) + (N - N_c - C_r) * \Theta(1) + C_r * \Theta(1)$$
$$= \Theta(N)$$

and $T_{CF} = T_1 + T_2 + T_3 + T_4 = \Theta(N)$. The above explained broadcasting of messages by different category of network nodes is summarized in Table 1.

Now, suppose $\Delta_{max}$ is the maximum number of neighbors of the nodes in the network, i.e., the maximum degree of the underlying network graph/topology, then each cluster may contain at the most $\Delta_{max}$ CMs and hence in the worst case, we have totally $N/(\Delta_{max} + 1)$ clusters/CHs and $N*(\Delta_{max} / (\Delta_{max} +1))$ CMs. Here, $\Delta_{max}$ can be a constant or a function of N. If it is constant, then $M_2=M_3=\Theta(N)$ .

Table 1. Summary of messages broadcast by network nodes for cluster formation

| Message | Broadcasting node | Time of broadcasting |
|---|---|---|
| Hello(my_id, my_pos) | All nodes | Periodically @ BI |
| WEIGHT_INFO(id, wt) | All nodes | Few seconds after start of simulation |
| CLUSTER_INFO(id, wt) | CH Critical nodes | After weight comparison |
| Cluster_id(id, my_cid) | CM | After joining to some cluster |

## 6.3 Overhead due to Cluster Maintenance

The maintenance phase is invoked in any of the four situations discussed in section 5.4. But, in all these cases, the nodes adjust themselves by sending either a *find_CH* message or a resignation message. The resignation message *CH_RESIGN(id)* is sent only by the CHs going beyond their battery limit and the message complexity for such transmission is $\Theta(\Delta_{max})$ . On the whole, in the worst case, such a message can be transmitted by all the CHs to their respective neighbors. But, this is equivalent to invoking the clustering procedure again. The *find_CH* message is broadcast in all the four cases.

Case i will lead to m number of broadcasts, where m denotes the number of CMs attached to the resigned CH. Case ii will lead to $n_a$ broadcasts, where $n_a$ denotes the number of newly added nodes, (one message corresponding to each newly added node). Case iii will cause two broadcasts, because any link failure will lead to change of status of two nodes joined by the link. Similarly, in Case iv, there will be two broadcasts for each newly added link. Thus, the total number of *find_CH* broadcasts is $m + n_a + (2*l_f) + (2*l_a)$, where $l_f$, $l_a$ denote respectively the number of link failures and link establishments.

## 7. RANKING OF CLUSTERS

In the cluster formation and adjustment procedures given in sections 5.3.1. and 5.3.2., the cluster-head is primarily elected based on the node weight and then the clusters are formed with the cluster-head and its neighbors. While generating clusters in this fashion, the neighbors of the cluster-head which are considered to be cluster members may be either strong neighbors, medium neighbors or weak neighbors. Based on this criterion, it is possible to rank the clusters

If all the cluster members in a particular cluster are strong neighbors of that cluster-head, then the connectivity in this case will be strong enough and the cluster-head can retain its status for a longer period. But, this will not cause the problem of quick battery drainage due to the strategy followed in assigning node weights. Thus, the clusters generated with the above features are ranked as the topmost (in terms of stability) clusters and are termed as ***Strong or Balanced Clusters***.

Similarly, the clusters whose cluster members are all medium neighbors are referred to as ***Medium or Semi-balanced Clusters***. The clusters whose cluster members are all weak neighbors are ranked as *Weak or **Unbalanced Clusters***.   There may be cases in which the cluster members are combinations of strong, medium and weak neighbors and such clusters are termed as ***Intermediate Clusters***.   In due course, because of the dynamic nature of our network, these Clusters can change into any of the above mentioned three types of clusters.

## 8. SIMULATION STUDY

### 8.1. Simulation Parameters

The validity of the theoretical analysis and the performance evaluation of SD_DWCA are obtained via simulation of the algorithm using NS-2 with CMU wireless extension.   We simulate a system of N nodes whose positions are randomly generated on a terrain of size XxY with a restricted diameter using the scenario generator available in NS-2 with the input parameters as listed in Table 2.

Table 2. Simulation Parameters for SD_DWCA

| Parameter | Meaning | Value(s) used in Simulation |
|---|---|---|
| N | Number of nodes | 50,75,100,150,200 |
| XxY | Terrain size/Network deployment area | 750x750 |
| Max Speed | Maximum Speed with which the nodes move in random directions | [0, $V_{max}$], where $V_{max}$ = 10, 20, 30, 40, 50m/s |
| Tx Range | Transmission range | 100, 150, 200, 250, 300 (metres) |
| PT | Pause Time | 30s |
| Mobility Model | The mobility model based on which the nodes move in random directions | Random Way Point mobility model |
| IE | Initial Energy | 100 Joules |
| Mac Type | Type of MAC layer | 802.11 |
| Traffic | Type of application used in data transmission between sources & destinations | CBR |
| Packet size_ | Size of packet transmitted | 1000bytes |
| Interval_ | Time interval between transmission of packets | 0.005s |
| T | Simulation time | 500s |

### 8.2. Performance Metrics

To measure the performance of our algorithm, the following metrics are considered.

1. Average number of Clusters – The number of clusters generated on an average during the simulation run time of our algorithm

2. Rate of re-affiliations – The number of cluster members which change their affiliations (Event of link breakage with the current cluster-head and link establishment with a new cluster-head due to mobility) per unit time.

3.  Average end-to-end throughput – The number of bytes successfully transmitted per unit time. i.e., the ratio of the transmitted packet size to the average end-to-end delay

4.  Overhead due to cluster formation – The number of control packets required to be transmitted for successful formation of clusters

5.  Overhead due to cluster Maintenace – The number of control packets to be transmitted to update the change in cluster structure due to mobility of nodes and arrival of new nodes

### 8.2.1. Average number of clusters

The following graphs are plotted to analyze of impact of transmission range and mobility on the average number of clusters for a varied number of nodes in the network. Figure 2 gives the impact of transmission range on the average number of clusters. The graphs are plotted for Transmission range varying from 100-300 (in steps of 50) for varied number of nodes ranging from 50 to 200, keeping maximum displacement fixed.

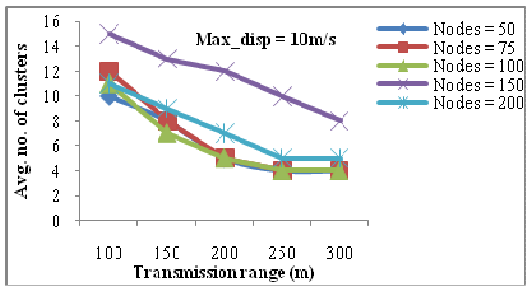
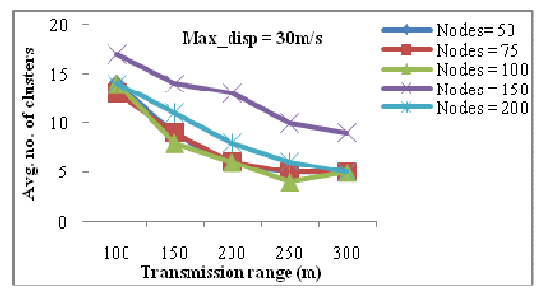

Figure 2(a) (Max_dip =10m/s)           Figure 2(b) (Max_dip =30m/s)

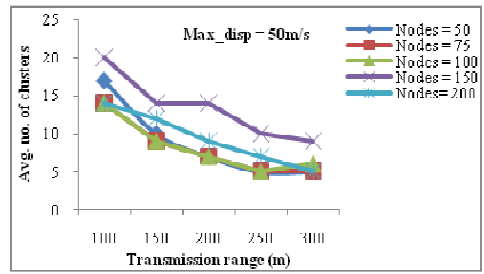

Figure 2(c) (Max_dip =50m/s)

Figure 2. Impact of Transmission range on average no. of clusters

It can be seen from the above graphs that the average number of clusters generated decreases as the transmission range increases, which is a natural phenomena. This is due to the well-known fact that as the communication range of each node increases, more number of nodes can be grouped into a single cluster thereby reducing the number of clusters generated.

Again, to analyze the impact of mobility, the graphs are plotted for Max speed varying from 10-50m/s (in steps of 10) for varied number of nodes ranging from 50 to 200, keeping transmission range fixed. This analysis is also performed to see the validity of the algorithm for high-speed networks. It can be witnessed from the graphs shown in Figure 3(a), (b), (c) & (d) that though the number of clusters generated increases with increase in velocity/maximum speed, the rate of increase is negligibly small. In some cases (Figure 3(b), (c)), the number of clusters obtained remains almost constant with increase in velocity. This guarantees that the algorithm can be used to cluster high-speed networks.

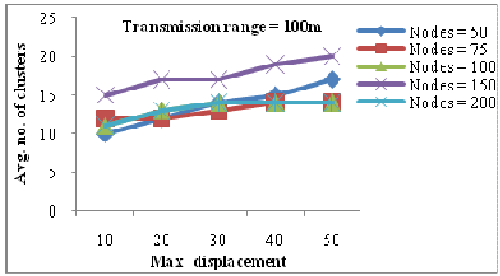

Figure 3(a) (Tx range =100m)

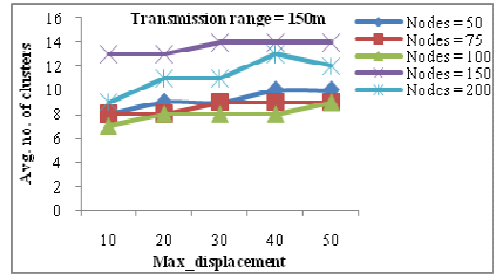

Figure 3(b) (Tx range =150m)

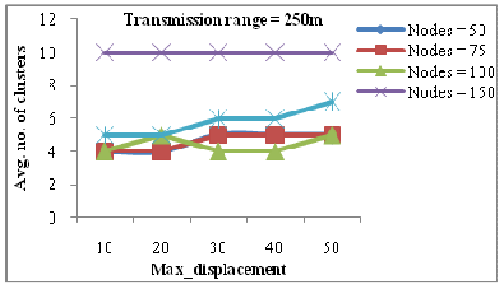

Figure 3(c) (Tx range =250m)

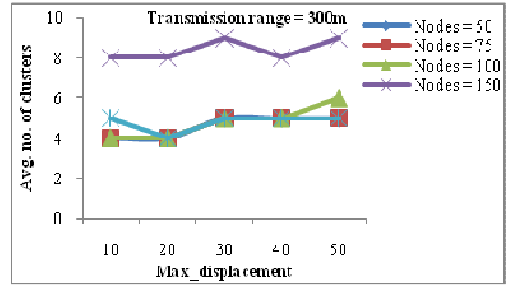

Figure 3(d) (Tx range =300m)

Figure 3. Impact of Mobility on average no. of clusters

### 8.2.2. Rate of re-affiliations

The following graphs are plotted to analyze the stability of the cluster structures developed. From the graphs shown in Figure 4, it can be inferred that the rate of re-affiliation increases as the number of nodes increases and reaches a peak value (which is nearly 0.02 much lesser than unity) and after that decreases to almost 0.005. The same behaviour is analyzed even with increase in maximum speed with which the nodes are allowed to move. This shows that the proposed algorithm generates highly stable clusters even for networks with high mobility. This is guaranteed by the choice of the weight parameters.

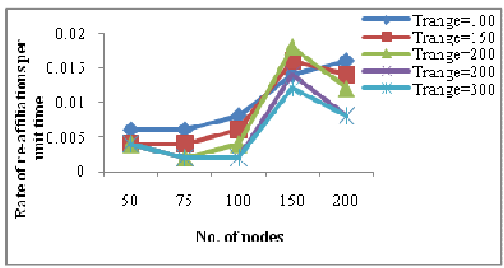

Figure 4(a) (Max_disp =10m/s)

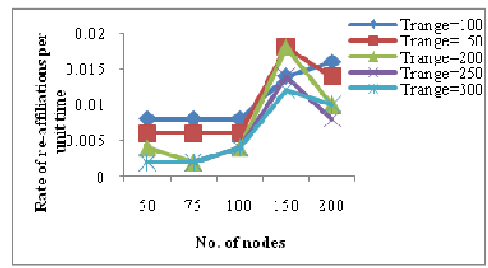

Figure 4(b) (Max_disp =20m/s)

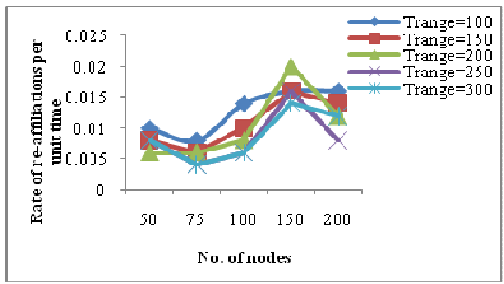

Figure 4(c) (Max_disp =40m/s)

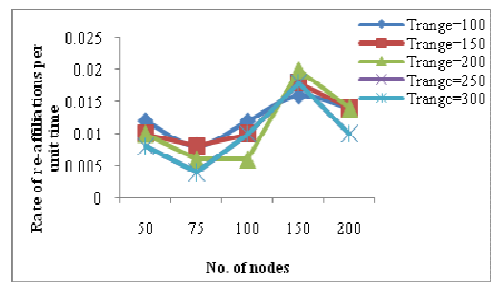

Figure 4(d) (Max_disp =50m/s)

Figure 4. No. of nodes *vs.* Rate of re-affiliations

### 8.2.3. Average end-to-end throuyput

Communication throughput influence the operation and performance of some applications. Hence, it is essential to investigate the throughput performance for ad hoc wireless networks. The average throughput is computed by dividing the packet size by the average end-to-end delay and the actual throughput is computed by dividing the total number of bytes received by the total end-to-end delay. [25]

In this paper, we have the average throughput as a performance metric obtained by dividing 1000 (Packet size used in the simulation) by the average end-to-end delay. It is measured in terms of kilobytes per second.

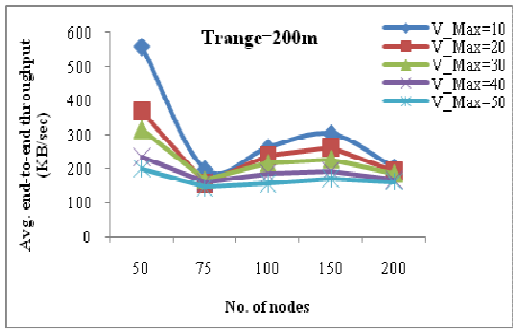
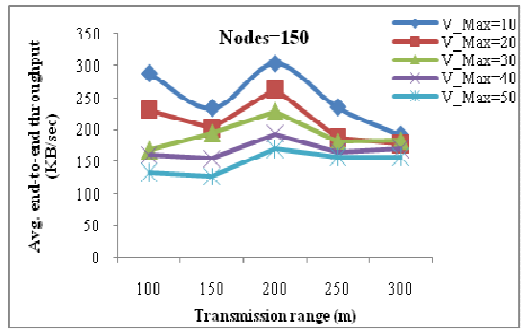

Figure 5. Nodes *vs.* Throughput          Figure 6. Tx Range *vs.* Throughput

The graph in Figure 5 is plotted by varying the number of nodes against throughput. As the number of nodes increases the throughput decreases and attains a saturation point and after that the throughput again increases, decreases and attains a saturation point. Hence, the throughput oscillates but varies between 150 to 400 KB/sec. Similar behavious is seen in the case of Figure 6, where the graph is plotted by varying the transmission range against throughput for by giving different maximum mobility speed.

### 8.2.4. Overhead due to Cluster set-up

As in theoretical analysis, the overhead incurred by the proposed algorithm is discussed under two cases. One is for initial cluster formation and the other for cluster maintenance. The algorithm is developed with the objective of avoiding re-clustering and minimizing the number of re-affiliations thereby reducing the overall cluster maintenance cost.

It can be seen from Figure 7(a), the overhead incurred is minimum in general and that is also maintained at a constant level even for increase in number of nodes and also increase in transmission range keeping maximum speed fixed. Similarly, from Figure 7(b), the overhead increases as the transmission range increases when the number of nodes is kept fixed.

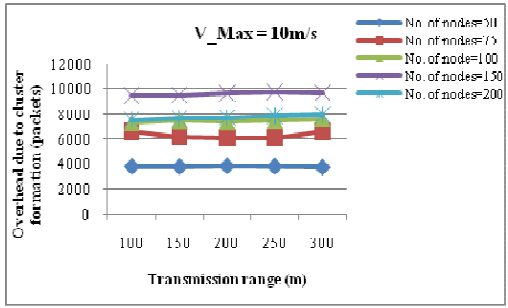

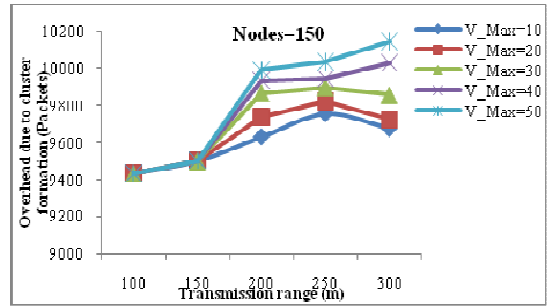

Figure 7(a). Tx range *vs.* Overhead
(keeping max speed fixed)

Figure 7(b). Tx Range *vs.* Overhead
(keeping no. of nodes fixed)

### 8.2.5. Overhead due to Cluster Maintenance

From the graph given Figure 8(a) and (b), it can be seen that the overhead due to cluster maintenance also has a similar behaviour, when plotted against transmission range and number of nodes keeping maximum speed at 10m/sand transmission range 200m fixed respectively.

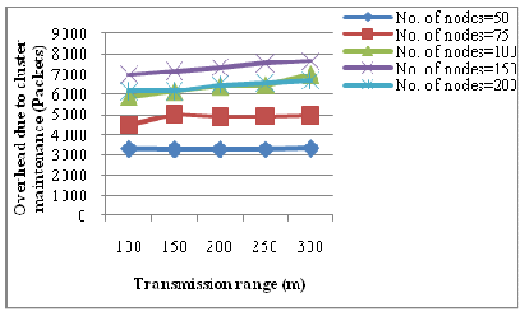

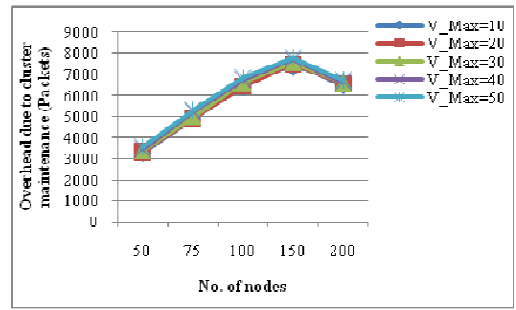

Figure 8(a). Tx range *vs.* Overhead
(keeping max speed=10m/s)

Figure 8(b). No. of nodes *vs.* Overhead
(keeping Tx range=200m)

As mentioned already, the avoidance of re-clustering will be meaningful provided the overhead due to cluster maintenance is considerably lesser than that of overhead of cluster formation. This is validated by the graphs shown in Figure 9(a) and (b). Here OH_set-up denotes overhead due to cluster set-up and OH_CM denotes overhead due to cluster maintenance.

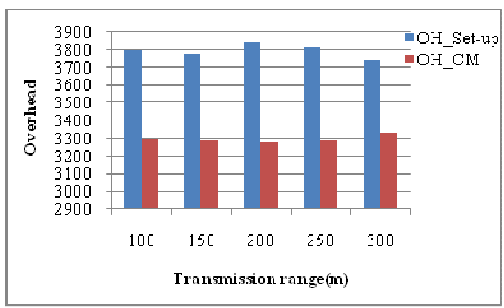

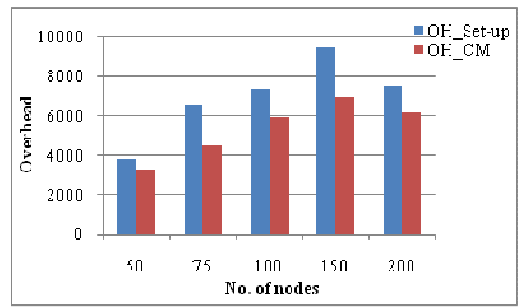

Figure 9(a) Tx Range *vs.* Overhead

Figure 9(b) No. of nodes *vs.* Overhead

Figure 9. Comparison of overhead due to cluster set-up and cluster maintenance

## 9. CONCLUSION

In the networks considered here, we allow mobility of the nodes in all possible directions. Hence, a node which is a weak neighbor at an instant may become a medium or strong neighbor at a later instant. Similarly, a medium neighbor may either become a strong or a weak neighbor and this leads to re-election of CHs. But, this situation can be avoided as much as possible by the choice of our mobility metric. Thus, the algorithm is reliable even for high speed networks. It is evident from the adjustment procedures given in section 5.3.2 that to reduce the number of critical nodes, we adopt re-affiliation rather than re-clustering and this greatly reduces the cost of cluster maintenance. Thus, the proposed algorithm is efficient in terms of both execution and maintenance. The experimental results obtained support the theoretical observations.

## Authors

**T.N. Janakiraman** is currently Associate Professor of Department of Mathematics, National Institute of Technology, Tiruchirapalli, India. He completed his undergraduate Studies at Madras University, India in 1980 and completed his Post graduation at National College, Trichy, India in 1983. He did his Ph.D. in Mathematics (Graph Theory and its applications) at Madras University with a UGC sponsored research fellowship and received his doctoral degree in the year 1991. He was a Postdoctoral Research associate for 1 year (1993-1994) in Madras University under the He has two sponsored research projects to his credit and published around 70 papers in refereed National/International journals. His research interests include Pure Graph Theory, Applications of Graph Theory to Fault tolerant networks, Central location problems, Clustering of wired & wireless ad hoc networks, Clustering of cellular and flexible manufacturing models, Image processing, Graph coding and Graph Algorithms.

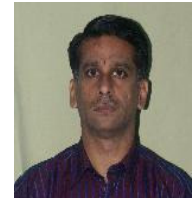

**A. Senthil Thilak** is currently a research Scholar of Department of Mathematics, National Institute of Technology, Tiruchirapalli, India. She received her Master's degree in Mathematics and Master of Philosophy in Mathematics from Seethalakshmi Ramaswami College, Tiruchirapalli, India. She has completed Post Graduate Diploma in Computer Applications in Bharathidasan University, Tiruchirapalli, India. She has published three papers in refereed National/International Journals. Her main research interests include Pure Graph Theory, Algorithmic Graph Theory and applications of graph theory to wireless ad hoc networks.

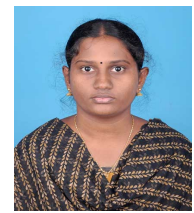